\documentclass[%
reprint,
superscriptaddress,
frontmatterverbose,
 amsmath,amssymb,
aps,
prb,
]{revtex4-1}

\usepackage{amsfonts,amsmath,amssymb,amsthm}

\usepackage{mathtools,siunitx} 

\usepackage{graphicx}
\usepackage{dcolumn}
\usepackage{bm}
\usepackage{titlesec}

\usepackage{hyperref}

\hypersetup{
	pdfnewwindow=true,      
	colorlinks=true,       
	linkcolor=blue,          
	citecolor=blue,        
	filecolor=blue,      
	urlcolor=blue,          
}

\usepackage{color}

\renewcommand\boldsymbol{\bm}

\usepackage{lineno}
\makeatother
\usepackage{ulem}

\newcommand{\be}{\begin{equation}}
\newcommand{\ee}{\end{equation}}

\newcommand{\ii}{{\mathrm{i}}}

\newcommand{\overbar}[1]{\mkern 1.5mu\overline{\mkern-1.5mu#1\mkern-1.5mu}\mkern 1.5mu}

\newcommand{\ncto}{{Na$_2$Co$_2$TeO$_6$}}
\newcommand{\ncso}{{Na$_3$Co$_2$SbO$_6$}}

\newif\ifdraft
\drafttrue
\draftfalse
\begin{document}

\preprint{APS/123-QED}

\title{Spin interaction and magnetism in cobaltate Kitaev candidate materials: an \textit{ab initio} and model Hamiltonian approach  }

\author{Shishir Kumar Pandey}
\affiliation{International Center for Quantum Materials, School of Physics, Peking University, Beijing 100871, China}

\author{Ji Feng}\email{jfeng11@pku.edu.cn}
\affiliation{International Center for Quantum Materials, School of Physics, Peking University, Beijing 100871, China}


%

\date{\today}

\begin{abstract}
In the quest for materials hosting Kitaev spin liquids, much of the efforts have been focused on the fourth- and fifth-row transition metal compounds, which are spin-orbit coupling assisted Mott 
insulators. 
Here, we study the structural and magnetic properties of 3$d$ transition metal oxides, Na$_2$Co$_2$TeO$_6$ and Na$_3$Co$_2$SbO$_6$. The partial occupancy of sodium in former compound is 
 addressed using a cluster expansion, and a honeycomb lattice of 
sodiums is found to be energetically favored. Starting from the \textit{ab initio} band structures, a many-body second order perturbation theory leads to a pseudospin-$\frac{1}{2}$ Hamiltonian
 with 
estimated magnetic interactions. We show that the experimentally observed zigzag magnetic state is stabilised 
only when the first neighbour Kitaev coupling dominates over the Heisenberg term, both of which are highly suppressed due to presence of $e_g$ orbitals. A third neighbour Heisenberg interaction 
is found dominant in both these compounds. We also present a phase diagram for Na$_2$Co$_2$TeO$_6$ by varying the electron-electron and spin-orbit interactions. The computed spin
 excitation spectra are found to capture essential features of recent experimental magnon spetrum, lending support to our results.

\end{abstract}

\maketitle

\section{Introduction} 
Strong quantum fluctuations in the ground state of a many-spin system can produce a quantum spin 
liquid state, in which spins are highly entangled even at large separations without any long-range order~\cite{bal, and}. 
Realisation of quantum spin liquid in actual materials is clearly attractive, especially for its potential application in quantum 
 computation~\cite{kitaev}, and high-$T_{\text c}$ superconductivity~\cite{and2}. 
 Known candidate materials that may host a spin liquid state 
are few~\cite{rmp1,rmp2,rev_jpcm,nat_rev}, and the quest for new potential candidates has recently drawn a lot of attention triggered by the pioneering work of  
Kitaev who proposed an exactly solvable spin model on triply coordinated lattice, exhibiting degenerate spin-liquid ground states~\cite{kitaev}.  
Extension of the Kitaev's honeycomb lattice model~\cite{khul1} to the inclusion of bond-dependent, highly anisotropic Heisenberg coupling interactions inherently introduces frustration to
 the spin system, and may be conducive to the realisation of quantum spin liquid states. 
 
A host of transition-metal compounds have been proposed as a material candidates for realising the Kitaev 
physics~\cite{co1,co2,co3,co4,co5,co9,ru1,ru2,ru3,ir1,ir2,ir3,ir4,ir5,ir6}.
Nevertheless, compounds containing a 3$d$ element distinguishes themselves from 4/5$d$ compounds.
 On the one hand, crystal fields (CF) and spin-orbit coupling (SOC) are believed to be substantially stronger in 4/5$d$ compounds than in their third-row counterparts. 
 On the other hand, 3$d$ elements are considerably more compact and the localisation of $d$ orbitals results in larger Hubbard $U$ and intra-atomic Hund's coupling compared to 4/5$d$ compounds.
  Much of  previous attention has been paid to 4/5$d$ magnetic materials, in which the interplay of Hubbard interaction ($U$), 
 intra-atomic Hund's exchange ($J_h$), CF
($\Delta$) and the SOC ($\lambda$)  leads to intricate Mott insulating states~\cite{mod1,mod2,mod3,mod4,mod5,mod6,mod7,mod8,mod9,mod10,mod11,
mod12,mod13}.
The studies focusing 3$d$ material are few and limited mostly to theoretical models~\cite{co1,co2,co4}. 
A quantitative examination based on \textit{ab initio} electronic structure methods of the electronic and magnetic structure of  the co-based compounds, \ncto~and 
\ncso , is evidently in order.

Each Co$^{+2}$ with seven $d$ electrons in a nearly octahedral CF of oxygens in these materials shows a high-spin $t_{2g}^5e_g^2$ configuration. 
The $d$ electrons on an isolated Co ion is then described by an effective spin $S = 3/2$ and orbital angular momentum $L=1$, giving 
rise to a low-energy pseudospin-1/2 doublet after application of SOC. Similar to iridates and RuCl$_3$, experiments propose zigzag magnetic ground state also for 
\ncto~and \ncso~\cite{co6,co7,co8}. 
What makes these $d^7$ cobaltates even more intriguing, compared to those  of  $d^{5}$ 
 iridates or RuCl$_3$, is the additional presence of spin-only active $e_g$ orbitals.

In transition metal compounds, the origin of anisotropic magnetic interactions such as Kitaev interaction is attributed to a combination of the directional 
nature of transition metal $d$ orbitals and the interaction of unquenched orbital moments with spin moments via spin-orbit interaction (SOI), 
which may give rise to this type of non-trivial anisotropic exchange coupling.   Under the large CF present in these 4/5$d$ compounds,  
the low energy excitations of $d^5$ configuration of transition metal ions can be described by a single hole with an effective spin moment 
$S  =1/2$ and an effective orbital angular moment $ L=1$ within the $t_{2g}$ manifold. The SOI then leads to an effective total angular momentum 
of $J_{\text{eff}} = S-L$,  
forming for a pseudospin-1/2 corresponding to doubly degenerate  ground state.


 Based on the above discussions, several interesting questions arises about structural, electronic and magnetic properties of cobaltates. 
 Whether cobaltates are a Kitaev-type material, and if so, how close it is to a quantum spin liquid, are questions worth pursuing. Additionally, whether the small structural difference in the two compunds, explained later, can bring some noticable changes in the magnetic interactions would be a crucial information from an experimental perspective.
Pertaining to the structure-property relation, how the Na partial occupancy in  \ncto~  impacts its electronic or the magnetic properties clearly requires investigation.

 In the present study, the fractional occupancy of Na in \ncto~is first resolved computationally using the cluster expansion technique, with which we obtain 
 particular pattern formation of Na atoms in the supercell of \ncto.  
Subsquently, the magnetic interactions between the Co$^{+2}$ ions in terms of pseudospins-1/2 for both \ncto~and~\ncso~are obtained based on the second-order perturbation theory. The hopping amplitudes and CF splitting considered in this method are estimated from tight-binding (TB) 
models obtained from fitting the $ab$ $initio$ electronic structures.  We show that 
 Heisenberg interactions as well as off-diagonal couplings are all highly suppressed and the dominant interaction is Kitaev-type for first nearest-neighbour 
 (1NN) Co atoms. Trigonal CF splitting, albeit comparable 
to the strength of SOC in these materials, does not alter the pseudospin-1/2 picture. The second nearest-neighbour (2NN) and inter-layer magnetic interactions are found to be negligibly small. 
Contrary to 4$d$/5$d$ based Kitaev-materials, quite surprisingly, the third nearest-neighbor (3NN) magnetic interactions are 
significantly larger than the 1NN ones. Remarkably, a suppressed Kitaev coupling and appearance of 
off-diagonal terms are revealed for 3NNs. The spin wave spectra obtained using these magnetic interactions in a Heisenberg-Kitaev 
model shows qualitative agreement with experiments. By varying parameters Hubbard $U$ and SOC strength-$\lambda$ we obtain 
a phase diagram for \ncto. We find different type of zigzag magnetic ordering stabilised in major portion of the phase diagram. 
We extensively discuss the properties of \ncto~in this study and finish our discussion in the end with a brief comparison between 
the properties of \ncto~and~\ncso.

\section{Methods}

Before proceeding to the discussions of the crystal and basic electronic structures, let us briefly describe the computational details of the \textit{ab initio} 
calculations for basic electronic 
structure and total energies, and of the cluster expansion and Monte Carlo simulation employed to understand
the Na vacancy in \ncto.

\subsection{ $Ab$ $initio$ and TB calculations}

To calculate the total energies and electronic structure of cobaltates, we perform density-functional theory (DFT)
calculations with the generalized-gradient approximation (GGA) within Perdew-Burke-Ernzerhof framework~\cite{PBE}. Calculations were performed using Vienna $ab$ $initio$
 simulation package (VASP: version-5.4.4)~\cite{Kresse}.
 Planewave basis set with a cutoff at 550 eV with projected augmented wave 
potentials~\cite{PAW,PAWpotentials1} is used, with a $8 \times 8 \times 4$/$6 \times 3 \times 6$ $\bm k$-grid for the Brillouin zone sum for calculations 
on the primitive cell of \ncto/\ncso, and a $6 \times 6\times 3$  $\bm k$-grid  for the supercell to be described later. To account for the correlation in Co(II) $d$-states, 
a static on-site $U$ is used in the GGA+$U$ formalism\cite{dudarev}. SOI was included along with spinor wavefunctions in the nonrelativistic approximation. The total energies are calculated self-consistently 
till the energy difference between successive steps was better 
than 0.5$^{-6}$ eV/formula unit. All results are sufficiently converged for this choice of basic computational parameters.
Using Na vacancy configuration obtained from our cluster expansion calculations (discussed next) along with the experimentally observed 
zigzag magnetic ground state,  
a full structural  optimization relaxing both ion positions and lattice parameters using $U$ = 3 eV leads to lattice parameters within 1.4\% of the experimental values. 
We hence use the experimental crystal structures in all subsequent $ab$ $initio$ band structure calculations.

To extract the CF matrix and hopping matrix elements, a TB Hamiltonian describing the low-energy states of both the materials were constructed using the maximally 
localized Wannier 
function implemented in Wannier90 package~\cite{wannier90}. We consider all the five Co $d$ orbitals in the TB basis. The strength of SOI $\lambda$ = 65 meV is determined by fitting 
the SOC included $ab$ $initio$ band structure with the TB model after explicitly introducing the intratomic SOC term ($\sum_i \lambda \bm L_i \cdot \bm s_i$ where $i$ is site index) in the model. 
The band structure fit is shown  in Fig. \ref{fig:fig03}(b).

\subsection{Cluster expansion and Monte Carlo simulations}

The issue of Na partial occupancy in \ncto~is resolved using a cluster-expansion model to aid a global survey of the 
 energy landscape of the vacancy disorder. The total energy of \ncto~with 2/3 Na occupancy in \ncto~is expressed as a cluster expansion up to the 4$^{th}$ order, as
 \begin{eqnarray}
 E(\mathcal C) = E_0 + \sum_{\alpha} g_{\alpha}V_\alpha \varphi_\alpha,
 \label{eq:cluster}
 \end{eqnarray} 
 where a configuration $\mathcal C$ refers to particular occupation of the lattice conforming to 2/3 occupancy of a Na monolayer, $V_\alpha$ are the effective cluster 
 interactions associated with the symmetry inequivalent cluster $\alpha$ with multiplicity $g_\alpha$, and 
 $\varphi_\alpha$ are the cluster correlations calculated as symmetrized averages of the products of all lattice sites (1, 2, 3 and 4 for one, 
 two, three and four-point clusters) 
 over the number of unit cells 
 needed to form the cluster $\alpha$ in  configuration $\{\mathcal C\}$.  
 Cluster spaces with up to four-point clusters (quadruplets) comprised of 132 distinct clusters are considered. 
 Further improvement on the model is not seen upon inclusion of higher order clusters. 
The effective cluster interactions  $V_\alpha$  are obtained using least absolute shrinkage and selection operator as implemented in the ICET package~\cite{icet}. 
The total energies of a
$2\times 2 \times1$ supercell obtained from \textit{ab initio} calculations are used to train and cross-validate the model in Eq.(\ref{eq:cluster}). 
   The model is subsequently employed in  Monte Carlo simulations in the canonical ensemble  to obtain the lowest energy configuration with Na vacancies. The Monte Carlo
  annealing processes start from a sufficiently high temperature of 1000 K, to avoid trapping into local minima.
  The annealing proceeds by lowering the temperature as a function of Monte Carlo steps with function $T_{\text{start}}$ - ($T_{\text{start}}$ - $T_{\text{stop}}$) $\times$ $\log$($step$ + 1) / $\log$($n_{\text{steps}}$) where 
  $T_{\text{start}}$ ($T_{\text{stop}}$) is the starting (target)  temperatures, and $n_{steps}$ ($step$) is the total number of Monte Carlo steps 
  (current step). We obtain the ground state 
  at 300 K in each case which remain 
so with further lowering of temperature close to zero. We set the lower temperature at 225
 K in our Monte Carlo simulations. 
The Na monolayer  $n\times n$ supercells with ($n= 4, 10, 14$) were used in our simulations. 
Typically, a system is annealed  in $2^6$ -- $2^8$ steps, depending on system size,  to avoid mode collapsing. 
MC simulations were performed using the same ICET package~\cite{icet}.

\section{ \texorpdfstring{\ncto}{TEXT}}

\subsection{Structure and Na vacancies}

\begin{figure}[ht]
  \centering
  \includegraphics[width=7.0 cm]{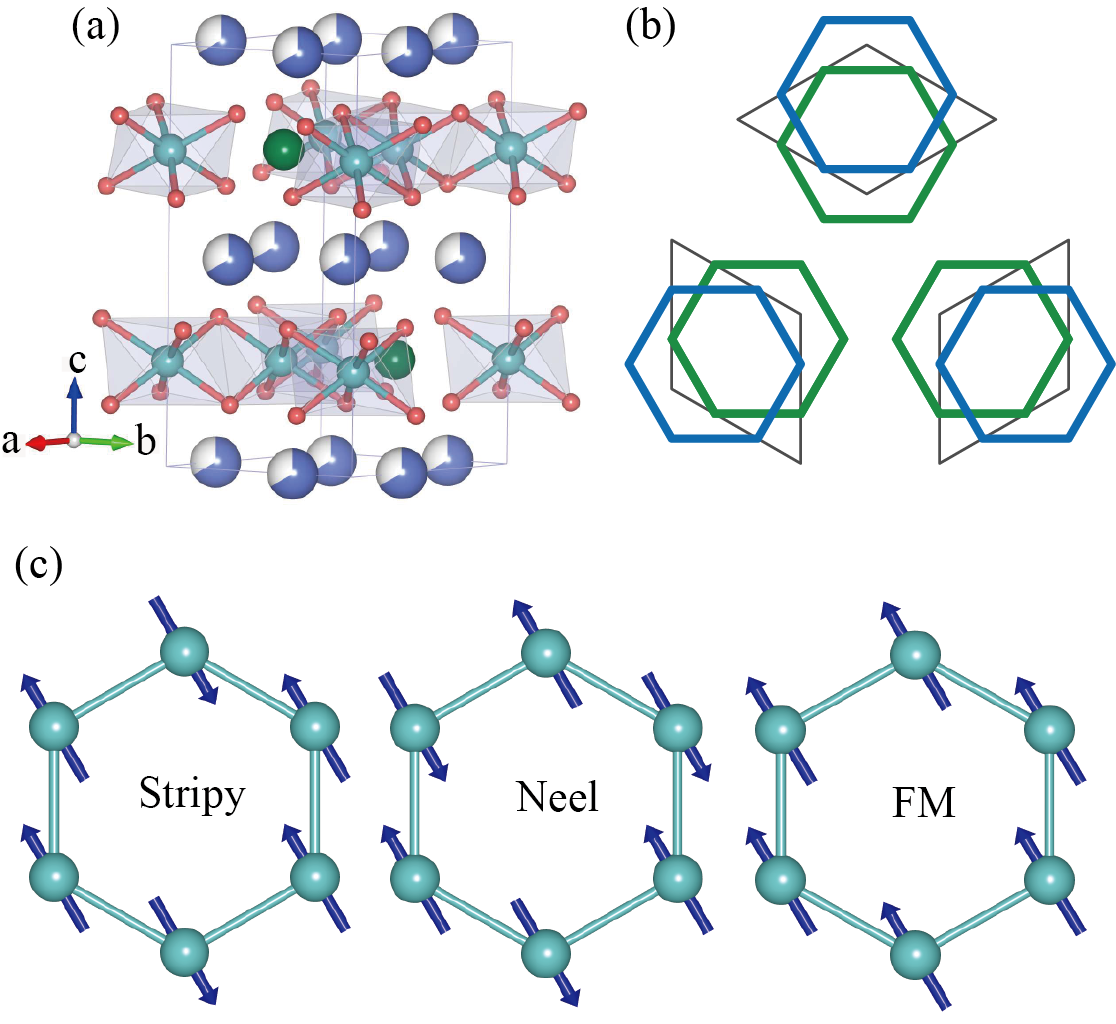}
  \caption{ Crystal and magnetic structures of \ncto. (a) Crystal structure.  Cyan, red and green balls represent Co, O and Te, respectively. 
Blue-white balls represent 
  2/3 partial occupancy of  Na atoms in \ncto.
  (b) Top view of the lowest-energy Na network in two layers of \ncto~unit cell obtained in our cluster expansion model, in accordance with an average 
  2/3 occupancy of the Na sites. Each of the layer 
  forms a honeycomb lattice, shown as blue and green hexagons, respectively. 
  The stacking of the bilayer breaks $C_3$ symmetry, as is evidenced by inspecting the three $C_3$ related replicas.
   (c) Possible magnetic structures in a single-layer honeycomb lattice of Co atoms, showing the stripy,  N\'eel and ferromagnetic configurations.}
    \label{fig:fig01}
  \end{figure}

In this section, we present the results concerning the basic structure and electronic structure of \ncto~obtained from  
\textit{ab initio} calculations. We will discuss properties of \ncso~later in Sec.~\ref{ncso}.
As shown in Fig.~\ref{fig:fig01}(a),  in the experimental crystal structure of  Na$_2$Co$_2$TeO$_6$~\cite{co3,co5,co6,co6,co7,co8,co9}  Co(II) ions are caged in edge-sharing oxygen 
octahedra, forming a honeycomb lattice with a Te ion sitting in the center of each hexagon.  These 2-dimensional Co-O-Te sheets are joined by  hexagonal close-packed Na monolayers to 
form a 3-dimensional array, with hexagonal lattice in the space group  $P6_322$.  It was found that the intercalated Na layers has a 2/3 overall occupancy, though the six-fold symmetry
 remains intact. Not only this presents a challenge to DFT calculations, the charge pattern of the partially occupied Na layers can also influence the spin states and interactions which
  is the central topic 
of present study. Therefore, we begin by sorting out this issue  before performing 
further analysis of the electronic and spin structures. 

\begin{figure}[ht]
    \centering
    \includegraphics[width=5.0cm]{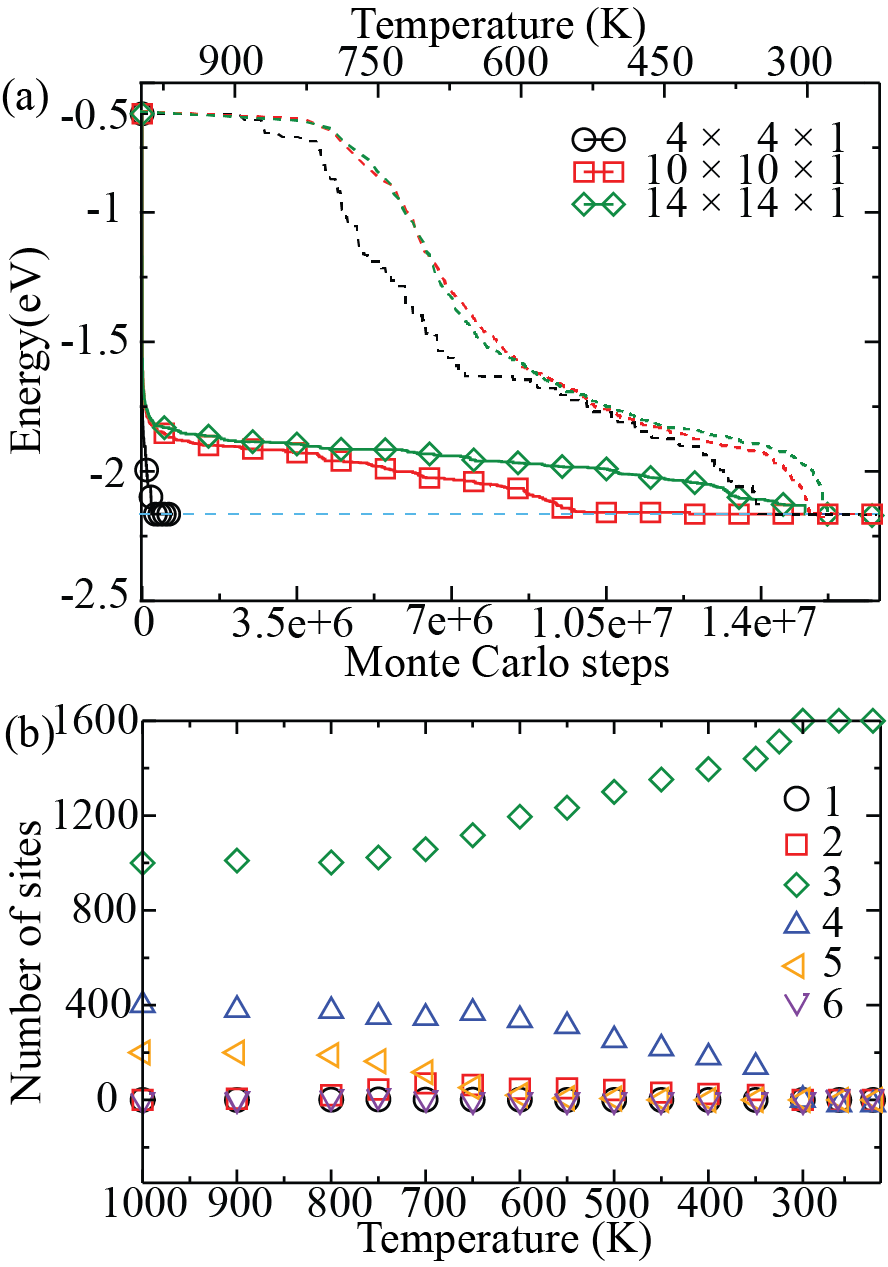}
    \caption{(a) Variation in configurational energy per atom with temperature (top three dashed lines)  and with Monte Carlo steps (lower three solid lines) obtained in annealing process for 
    various supercells. Horizontal dashed line represents the ground state energy. Different plateaus of energy in the equilibration 
    process (dashed lines) is clearly visible with the flat plateau in range 300-225 K revealing that ground state has reached. (b) Change in coordination of Na-sites with temperature for 10 $\times$10$\times$1 supercell. 
    For ground state, only 3-point clusters are present 
    with population of all other clusters vanishing with lowering of temperature.}
      \label{fig:fig02}
    \end{figure}

To this end, we perform cluster expansion calculations in which random starting vacancy configurations conforming to 2/3 occupancy invariably converge to a honeycomb lattice, 
when annealed  from over 1000 K steadily down to 300 K.  The energy evolution for different supercells in the simulated annealing processes is shown  in Fig.~\ref{fig:fig02}(a), 
which all evidently converge to the same ground state. We stress that the energy in this plot is the configurational potential energy of Na vacancies, which shows 
several steps in various temperature ranges before plateauing at 300 K for ground state configuration. Lowering the temperature further to 225 K does not change the energy, reassuring that the 
system is well equilibrated.
The evolution of  coordination numbers of Na for a 10 $\times$ 10 supercell during a typical annealing is shown in Fig.~\ref{fig:fig02}(b).
 The population of triply coordinated Na-sites increases gradually as the annealing proceeds,  while the population of atoms with other coordination numbers diminish. The resultant structure 
 at 300 K only has triply coordinated sites, corresponding to a honeycomb lattice.  These results point to  the honeycomb net as a most probable ground state monolayer configuration. However,
  when we stack two of such these honeycomb lattices together in a given experimental unit cell, where the original hexagonal close-packed sheets adopt a $ABAB$ type periodic stacking, 
  it is seen the unit cell cannot have 3-fold rotational symmetry. As shown in  Fig. \ref{fig:fig01}(b), the three replica of such bilayer stacking with a unit cell interrelated by $C_3$ 
  rotations are shown.
   Clearly, these replica do not coincide with each other, demonstrating the the lack of 3-fold symmetry in this \emph{ordered} Na vacancy configuration. 

We thus come to the following interpretation of the computed Na vacancy configuration, in order to reconcile with the experimentally established crystal structure with an apparent 
3-fold symmetry.  Within each layer, the Na vacancies are arranged to produce an ordered honeycomb lattice, and the layer stacking is disordered  owning to weak interlayer interactions
 to produce an overall 3-fold symmetry in elastic scattering experiments. For example, for an in-plane reciprocal lattice vector $\bm g \cdot \bm c =0$, the structure factor of the Na 
 sublattice is
$$
S({\bm g}) =  n \,  \overbar {\exp(-\ii \bm g \cdot \bm d)},
$$
where $n$ is the atom number density, $\bm d$ is the location of sodium in a 2-dimensional honeycomb sheet within a unit cell, and the overbar means averaging over all layers.  
Clearly, given the presumed stacking disorder, the three configurations in Fig. \ref{fig:fig01}(b) contribute equally to the average and $S(\bm g)$ shall have 3-fold symmetry. 

\subsection{Electronic structure}

Therefore, we shall adopt one of the three equivalent Na configurations in Fig. \ref{fig:fig01}(b) in the \textit{ab initio} calculations. Note, however, that the 3-fold symmetry will be 
imposed on the TB Hamiltonian  based on which magnetic interactions are extracted and discussed later. 

Once the SOI is included self-consistently in our calculations in the structure described above, the experimentally proposed zigzag antiferromagnetic ground state can be reproduced. 
The total energy for various magnetic configuration is listed in 
Table~\ref{tab:tot_en} and shown in Fig.~\ref{fig:fig01}(c) (For zig-zag configuration, see Fig.~\ref{fig:fig04}(b)). The spin magnetic moments of Co 
 atoms in the ground state were found to be 2.69 $\mu_B$, comparable with the experimentally observed value of 2.77 $\mu_B$ and a little smaller than 
 what is expected for an effective 3/2-spin system. 
 
 \begin{table}[h!]
 \centering
\begin{tabular}{lllll}
\hline{}
  &\multicolumn{3}{c}{Spin-direction} \\
\cline{2-4}
  Mag. config.   & 010   & 001 & 100 \\
\hline
 Zig-zag & 0.0     & 0.226  &   0.231     \\
  Stripy & 0.338        & 1.688      &  0.116      \\
Neel & 0.724      & 0.737       & 0.758 \\{}
FM & 0.252       & 0.073      & 0.086 \\
\hline 
\end{tabular}
\caption{Energies of various magnetic configurations (in eV) relative to the zigzag antiferromagnetic configuration.}
\label{tab:tot_en}
\end{table}

\begin{figure}[ht]
\centering
\includegraphics[width=7cm]{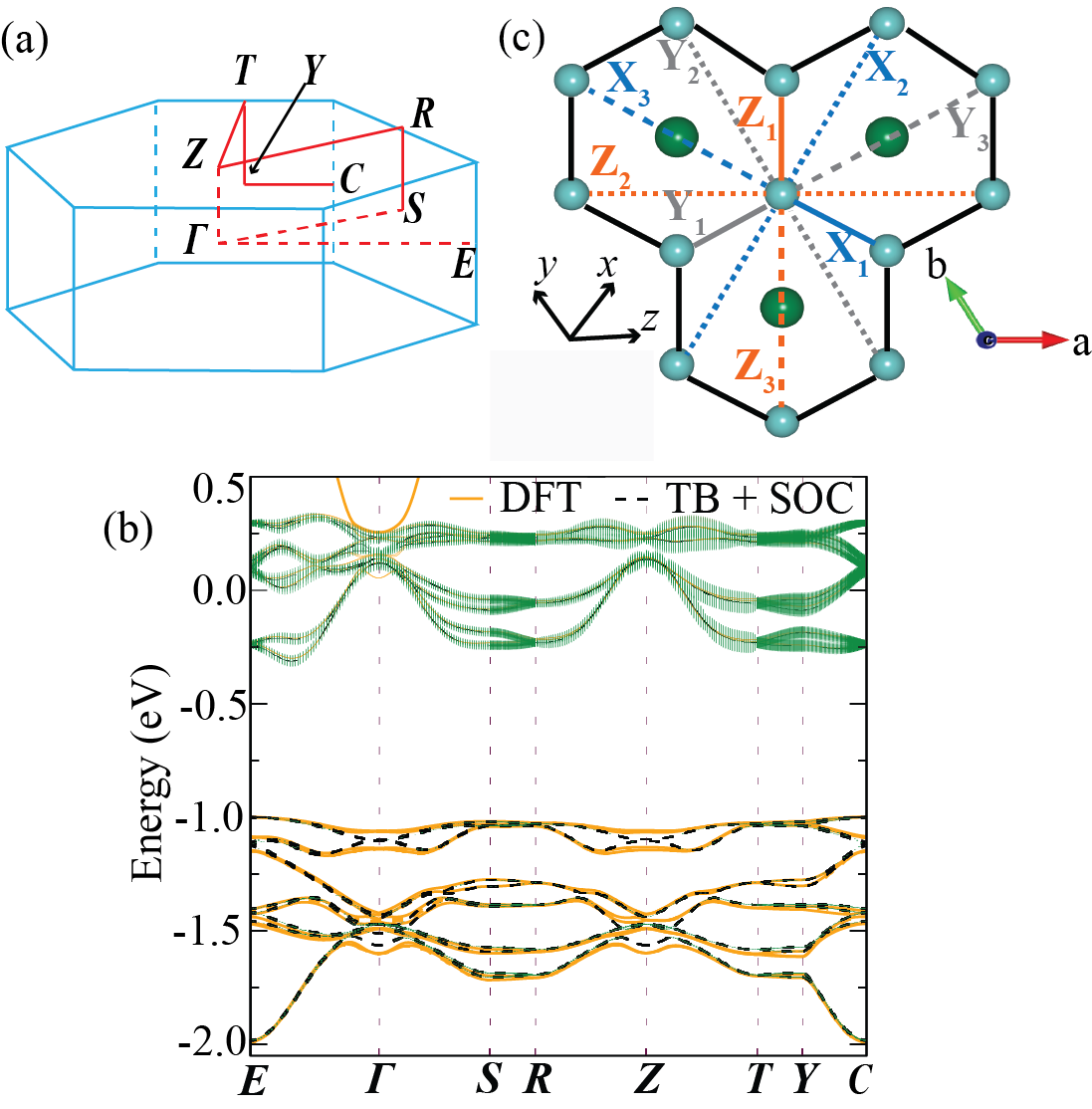}
\caption{(a) Brillouin zone of \ncto. (b) $Ab$ $initio$ band structure with SOC fitted with TB Hamiltonian with SOI.
The fat band representation depicted in green shows  $e_g$ contribution of Co-$d$ orbitals.
The Fermi energy is set to zero. (c) A single-layer Co$^{+2}$ lattice in \ncto. 1NN, 2NN and 3NN bonds
are indicated by solid, dotted and dashed lines, respectively. Orange, dark gray and skyblue  color represent the Z, Y and X bonds, respectively. $a$, 
$b$ and $c$ are the crystallographic axes and cubic axes of Co-O octahedra are denoted by $x$, $y$ and $z$.}
  \label{fig:fig03}
\end{figure}

In order to extract the magnetic interactions, we constructed a TB Hamiltonian as described in the methods section. The crystal structure of cobaltates have 
a $C_3$ axis along crystallographic $c$-axis. 
We choose the local axes $x$, $y$, $z$ along Co-O bonds such that $x$ + $y$ + $z$ = $c$, i.e. the local axes respects 3-fold symmetry of 
the crystal structure.
Mapping in this setting of local axes was done considering  a basis formed by all the five $d$ orbitals on Co atoms. The Brillouin zone of \ncto~is shown in  Fig.~\ref{fig:fig03}(a). The 
\textit{ab initio} band structure for the non-magnetic phase shown in Fig.~\ref{fig:fig03}(b) in a calculation involving SOC shows that the low-energy excitation involves both the
 $e_g$ and $t_{2g}$ states. Thus the Wannier functions involve the full set of $d$ orbitals on Co. Using the basis 
$\psi^{\dagger}_{i\sigma}$ = $[d^\dagger_{z^2}$,$d^\dagger_{xz}$, $d^\dagger_{yz}$, $d^\dagger_{x^2-y^2}$, $d^\dagger_{xy}]_{i\sigma}$ for site $i$ and spin $\sigma$, the TB Hamiltonian
 then reads
\begin{eqnarray}
H_0 &=& H_{\text{cf}} + H_{\text{hop}} + H_{\text{soc}} \nonumber \\
&=& \sum_{i,\sigma} \psi_{i\sigma}^\dagger {\Delta_i}\psi_{i\sigma} + \sum_{i\neq j,\sigma} \psi_{i\sigma}^\dagger T_{ij}\psi_{j\sigma} \nonumber \\
&+& \sum_i \lambda \bm L_i \cdot \bm s_i
\label{eq:h0}
\end{eqnarray}
in which the CF ($\Delta_i$) and the hopping amplitudes ($T_{ij}$) 
are obtained from the TB model without SOI.
The so-determined spin-independent CF matrix (in eV) is 
\begin{equation}
\label{eq:cf}
\Delta_i = 
\begin{bmatrix*}[r]
1.377  & 0.045 & 0.045 & 0.000 &  0.089 \\
0.045 &  0.015 & 0.006 & 0.078  & -0.006 \\	
0.045 &  0.006 & 0.051 &  -0.077  & -0.006 \\
0.000 & 0.078 & -0.078 &  1.377 &  0.000 \\
0.089 & -0.006 & -0.006 &  0.000 &  0.015 \\
\end{bmatrix*}
\end{equation}
This CF matrix has explicit $C_3$ symmetry. The spin-independent hopping amplitudes are listed in~\ref{appendix:hop}. These parameters are used in the 
next set of calculations.  
The last term in Eq.~(\ref{eq:h0}) is the atomic SOI. It is observed that the strength of SOI  ($\lambda$ = 65 meV, see Methods section) in 
\ncto~is considerably smaller than that 
of Ir- and Ru-based materials, for which $\lambda\approx$ 400 meV~\cite{ir7} and 150 meV~\cite{ru3}, respectively.

\subsection{Magnetic interaction and magnetism}

Before investigating in detail the magnetic properties of \ncto, it is important to discuss its structural and electronic features. A comparison between \ncto~and previously 
studied 4/5$d$ materials with pseudospin-1/2 ground state highlights following crucial differences between these materials. Additional presence of $e_g$ orbitals for Co$^{+2}$-$d^7$ 
with active spin only degree of freedom 
put this material in a different class when compared to the materials with $d^5$ configuration. 
Also, magnitude of trigonal CF in \ncto~($\sim$ 27 meV, see the matrix in Eq.~[\ref{eq:cf}]) is expected to be of the same order 
as that of the SOC strength ($\lambda$) which further distinguishes this material from those of 4$d$/5$d$ based compounds where 
 $\lambda$ is an order of magnitude larger than the trigonal CF. 
As $\lambda$ is responsible for energy separation between $J_{\text{\text{eff}}}$-1/2 and  $J_{\text{eff}}$-3/2 states 
while trigonal CF mixes these states, this separation is expected to be smaller in \ncto.

From a structural perspective, Co-O-Co angles across a pair of edge-sharing octahedra are $\sim$ 92$^\circ$ which is close to the ideal case of $90^\circ$. 
In contrast, the same metal-ligand-metal bond angles in  Na$_2$IrO$_3$ and RuCl$_3$ are significantly larger:
100$^\circ$ for the former and 94$^\circ$ for the latter~\cite{ir4,ru2}. 
This difference manifests itself in terms of enhanced direct ($\sigma$-type) hopping amplitudes between $d$ orbitals of Co atoms~\cite{khomski}. 
For example,
in~\ref{appendix:hop}, $d_{yz}-d_{yz}$ hopping on a X bond is larger when compared to other hopping interactions. The same is true for 
$d_{zx}-d_{zx}$  and $d_{xy}-d_{xy}$ hopping on Y and Z bonds.  Opposite has been observed in case of Na$_2$IrO$_3$~\cite{khomski}.

To be concrete, let's consider the interaction Hamiltonian
\begin{eqnarray}
H_{\text{int}} &=& \frac{U}{2} \displaystyle\sum_{i,\alpha} n_{i\alpha \sigma} n_{i\alpha \sigma'} 
+ \frac{U'}{2} \displaystyle\sum_{i,\alpha \ne \beta} n_{i \alpha}n_{i \beta} \nonumber\\
 \nonumber &-&\frac{J_\text{H}}{2} \displaystyle\sum_{i, \sigma, \sigma', \alpha \ne \beta} \psi^\dagger_{i\alpha \sigma} \psi_{i\alpha \sigma'} \psi^\dagger_{i\beta \sigma'} \psi_{i\beta \sigma} \\ 
 &-& \frac{J'}{2} \displaystyle\sum_{i, \sigma \ne \sigma', \alpha \ne \beta} \psi^\dagger_{i\alpha \sigma} \psi_{i\beta \sigma'} \psi^\dagger_{i\alpha \sigma'} \psi_{i\beta \sigma} 
\end{eqnarray}
Here,  $U$/$U'$ are intraorbital/interorbital Hartree energies;  and  $J_\text{H}$ and $J'$ are Hund's coupling and pair hopping interaction,
respectively. Rotational invariance in the single atom limit dictates the relationships:  $U'$ = $U$ - 2$J_\text{H}$ and $J_{\text H}$ = $J'$.

The key here is the $d^7$ manifold has a two-fold degenerate ground state, which form a  Kramers doublet and can be treated as a pseudospin-1/2.  In order to extract the magnetic 
interactions of these pseudospin states,  we first project the full TB Hamiltonian onto the pseudospin $J_{1/2}$ 
space $\{\phi_{i\alpha}\}$, $\alpha = \uparrow, \downarrow$, where $\uparrow,\downarrow$ refer to the SOC pseudospin-1/2 states. 
We note here that due to consideration of local axes $x$, $y$ and $z$ in the forthcoming calculations,
 up/down axis of the pseudo-spins are along $c$ axis which is also the trigonal distortion axis of  Co-O octahedra in \ncto. We start in the isolated atom limit, 
where $H_{\text{atom}} = H_{\text{cf}}+H_{\text{soc}}+H_{\text{int}}$, and introduce the hopping contribution $H_{\text{hop}}$ as a perturbation.  In the second-order perturbation theory, 
the Hamiltonian is written as,

\begin{widetext}
\begin{eqnarray}
H^{(2)} &=& \sum_{ij}\sum_{\alpha\beta\alpha'\beta'} \mathcal H (i,j)_{\alpha\beta\alpha'\beta'}
| i\alpha,j\beta\rangle\, \langle i\alpha', j\beta '|, \nonumber\\
\end{eqnarray}

\begin{eqnarray}
\mathcal H (i,j)_{\alpha\beta\alpha'\beta'} &=&
\sum_{kl}\sum_{\gamma\lambda}
\frac{1 }{\Delta E}
\langle i\alpha,j\beta|H_{\text{hop}}|k\gamma, l\lambda\rangle\,
\langle k\gamma,l\lambda|H_{\text{hop}}|i\alpha', j\beta'\rangle ,
\label{eq:h02}
\end{eqnarray}
\end{widetext}
where 
$1/\Delta E = \frac{1}{2}[{1}/(E_{i\alpha}+E_{j\beta} - E_{k\lambda}-E_{l\gamma})
+
{1}/(E_{i\alpha'}+E_{j\beta'} - E_{k\lambda}-E_{l\gamma})]$.
Here, $| i\alpha,j\beta\rangle$ and $| i\alpha', j\beta'\rangle$ are two-site states in the $J_{1/2}$ ground states, and $| k\lambda, l\gamma\rangle$ are two-site excited states, 
both in the isolated atom limit.  $H_{\text{hop}}$ connects a two-site ground state to an excited state with $(d^6, d^8)$ configuration, the dimensions of whose Hilbert spaces 
are 210 and 45, respectively.  The eigenstates of isolated Co with 6, 7 and 8 $d$ electrons are obtained by exact diagonalization. 

Considering the scenario when magnitude of Hubbard $U$ and CF splitting are much larger than the hopping amplitudes and also when the 
SOC strength is much larger than $t^2/U$, the low energy space is formed by the lowest two degenerate many body states of the ``onsite" Hamiltonian ($H_{\text{atom}}$). 
These states behaves exactly as psuedospin-1/2 Kramers doublet in the limit of an infinite CF and in the absence of any lattice distortions which can further 
split the $t_{2g}$ orbitals such as trigonal distortions.  Writing the pseudospin $J_{1/2}$ as $S^\mu = \frac{1}{2}\sigma^\mu$, 
Eq.~(\ref{eq:h02}) can be mapped to a spin Hamiltonian of the form
\begin{eqnarray}
H_{\text{spin}} &=& S_i^\mu \Gamma(i,j)^{\mu\nu} S_j^\nu \nonumber \\
&=& \frac{1}{4} 
\Gamma(i,j)^{\mu\nu}  \phi _{i\alpha }^{\dagger }\sigma _{\alpha\alpha '}^\mu \phi _{i \alpha '} 
 \phi _{j\beta } \sigma _{\beta\beta '}^{\nu}  \phi _{j\beta }^{\dagger },
\end{eqnarray}
where $\mu,\nu=0,x,y,z$, $\sigma^\mu$ are Pauli matrices, and summation over all repeated indexes is implied. The map can be achieved by solving the linear equations
\begin{equation}
-\frac{1}{4} \sigma_{\alpha\alpha'}^\mu \sigma_{\beta\beta'}^\nu 
\Gamma(ij)^{\mu\nu}
=\mathcal H(i,j)_{\alpha\beta\alpha'\beta'}
\end{equation}

The spin Hamiltonian thus can  be rewritten as, 
\begin{eqnarray}
\label{eqn:jk}
H_{\text{spin}} &=&  \sum_{\langle i,j \rangle \in l(mn)} [ J^l_1 \boldsymbol{\sigma}_i \cdot 
	            \boldsymbol{\sigma}_j  + K^l_1 \sigma_i^l \sigma_j^l \\ \nonumber 
	            &+& \eta^l_1 \left( \sigma_i^m \sigma_j^n + \sigma_i^n \sigma_j^m \right) \\ \nonumber
&+& \eta'^l_1 \left( \sigma_i^m \sigma_j^l + \sigma_i^n \sigma_j^l + \sigma_i^l \sigma_j^m + \sigma_i^l 
\sigma_j^n\right) ] \\ \nonumber
         &+& \sum_{\langle \langle \langle i,j \rangle \rangle \rangle \in l(mn)} [J^l_3 \boldsymbol{\sigma}_i \cdot 
	            \boldsymbol{\sigma}_j  + K^l_3 \sigma_i^l \sigma_j^l \\ \nonumber 
	            &+& \eta^l_3 \left( \sigma_i^m \sigma_j^n + \sigma_i^n \sigma_j^m \right) \\ \nonumber
&+& \eta'^l_3 \left( \sigma_i^m \sigma_j^l + \sigma_i^n \sigma_j^l + \sigma_i^l \sigma_j^m + \sigma_i^l 
\sigma_j^n\right) ]
\end{eqnarray}
Above $\langle i,j \rangle$/$\langle \langle \langle i,j \rangle \rangle \rangle$ represent 1NN/3NN pairs.  The 2NN interaction is omitted as they are negligibly small.
$J$, $K$ and $\eta$ are, respectively, the Heisenberg,
Kitaev and off-diagonal interactions on any $l\in$ Z/X/Y bond. E.g., for Z-type bond, ($mn$) is $(xy)$ and so on. 

We now comment briefly 
on the results obtained from diagonalization of the ``onsite'' Hamiltonian, $H_{\text{atom}} =H_{\text{cf}} + H_{\text{soc}} +H_{U}$, 
before going into a detailed discussion of the magnetic interactions in \ncto. 

\subsubsection{Onsite Hamiltonian: \texorpdfstring{$H_{\text{atom}}$}{hatom}}

There are four parameters in $H_{\text{atom}}$ viz, $U$, $J_{\text{H}}$, $\lambda$ and $\Delta$. 
From the optical spectra on CoO~\cite{coo}, estimated value of Hund's splitting ($J_{\text{H}}$) is $\sim$ 0.8 eV. 
$J_{\text{H}}/U$ ratio in cobaltates is believed to be $<$ 0.2~\cite{co1}, hence $U$ = 5 eV was initially fixed in our calculation 
of magnetic interactions.
 $\lambda$ = 65 meV obtained from band structure fitting is slightly 
larger than an estimated value of $\sim$ 0.015 eV (corresponding to $J_{\text{eff}}$-1/2 to $J_{\text{eff}}$-3/2 transition) from inelastic 
neutron scattering experiment~\cite{co3}. Later, we will vary both $U$ and $\lambda$ to examine their effect on the magnetic interactions.
$\Delta$ obtained from the atomic Hamiltonian of the TB model is consistent with what is expected for such 
materials~\cite{co1,co4}.

The energy separation between $J_{\text{eff}}$-1/2 and $J_{\text{eff}}$-3/2 states, obtained after diagonalizing $H_{\text{atom}}$ with above mentioned parameter values,  
is $\sim$ 24 meV which is in close agreement with the experimentally observed value of $\sim$ 21 meV in \ncto~\cite{co3}, 
 and is much smaller than the expected value of $\sim$ 100 meV corresponding to $\frac{3}{2}\lambda$ (for $\lambda$=0.065 eV). 
We attribute this difference to be arising from large mixing between $t_{2g}$-$e_g$ orbitals through the trigonal 
distortions present in \ncto. Compression of Co-O octahedra along $c$-axis causes large deviation of O-Co-O angles from the ideal 
90$^o$ with the smallest and largest angles being $\sim$ 78$^o$ and 97$^o$, respectively. This non-orthogonal local environment causes mixing 
between different $d$ orbitals. It is worth mentioning here that estimate of $\lambda$ from experiments would yield a value of 15 meV and yet 
we obtained $J_{\text{eff}}$=1/2 - 3/2 separation consistent with experimentally observed value using $\lambda$= 65 meV. This calls for scrupulous 
attention while estimating $\lambda$ in the systems with active $e_g$ orbitals and large trigonal distortions. 
The $J_{\text{eff}}$-3/2 states were found to split into two doublets with an energy separation of $\sim$ 7 meV. 
To examine the effect of ``imperfect'' CF due to additional trigonal distortions, we compare the result to a perfect octahedral geometry 
 with $\Delta$ = 1.36 eV. In this case, $J_{eff}$-1/2 and $J_{eff}$-3/2 separation increased to 34 meV while the $J_{eff}$-3/2 states regained their four fold degeneracy.  
 Although the CF heavily mixes $J_{eff}$=1/2-3/2 states, pseudospin-1/2 picture is still 
 relevant in \ncto~as both, the $U$ and CF splitting $\Delta_{t_{{2g}-e_g}}$ are much larger than the hopping amplitudes ($t$) and $\lambda \gg t^2/U$ 
 still holds in our case. 

\subsubsection{Magnetic interactions}

The  magnetic interaction on X/Y/Z bonds (see Fig.~\ref{fig:fig03}(c)), estimated from the process described above, are tabulated in Table~\ref{tab:m_i_u5}. 
Owing to the local site symmetry $C_3$, we have identical magnetic interactions 
on all the three bonds. Indeed, the 1NN Heisenberg term is highly suppressed and antiferromagnetic as was expected 
in this case due to opposite sign of $e_g$-$e_g$ and $t_{2g}$-$e_g$ exchange process~\cite{co1,co4}. However, contrary to 
the previous speculations, the ferromagnetic Kitaev term is also small and found to be an order of magnitude smaller 
when compared with recent experimental estimation from the fitting of spin-wave dispersion data~\cite{co3}. 
The same is true for a comparison with Ru~\cite{mod12} and Ir~\cite{mod11} based compounds. There can be mainly two reasons behind such a reduction. 
First, smaller hopping amplitudes and large $U$ in our case when compared to 4$d$/5$d$ materials can contribute to this huge 
 reduction. The effect of the latter is verified by reducing $U$ to 2.5 eV while keeping the $J_\text{H}/U$ ratio fix at 0.16, in which case the magnitude 
 of both $J_1$ and $K_1$ are substantially enhanced,  to 1.93 and $-1.616$ meV, respectively. The other factor contributing to the reduction is  the weaker SOI in \ncto.  We will examine 
 the effect of varying $\lambda$ in a later section to understand how it affects 
 the magnetic interactions.

\begin{table}[!ht]
 \centering
\begin{tabular}{lSSSSSSSS}
\hline
  &\multicolumn{3}{c}{First NN}  & & \multicolumn{3}{c}{Third NN} & \\
\cline{2-4} \cline{6-9}
  Bond type   & $J$   & $K$ & $\eta$/$\eta'$ & & $J$ & $K$ & $\eta$/$\eta'$ \\
\hline
X/Y/Z &  0.261    & -0.678  &   0.0 & & 3.153 &  -0.04 & $|$0.76$|$    \\

\hline 
\end{tabular}
\caption{ Estimated 1NN and 3NN Heisenberg $J$, Kitaev $K$ and off-diagonal $\eta$, $\eta'$ terms in \ncto~given in meV. The 2NN interactions as well as 
terms other than $J$ and $K$ for 1NN were found to be negligibly small. Parameters used are $U$ = 5 eV, 
$J_{\text{H}}$ = 0.8 eV and $\lambda$ = 65 meV.}
\label{tab:m_i_u5}
\end{table}

 Interestingly, all other first-neighbour off-diagonal spin interactions $\eta_1$ and $\eta'_1$ are found to be negligibly small for this particular set 
  of parameters, in contrast to what was obtained for Ir-~\cite{mod11} and Ru- based~\cite{mod12} compounds. Previous arguments for the 
  suppression of the Heisenberg interaction in this material also apply to these off-diagonal interactions.
We find the second-neighbour as well as inter-layer coupling to be negligibly small, ruling out the possibility of any inter-layer interactions and effect of 
Na-vacancies which separates these Co-hexagons in $c$-direction. This again is in complete agreement with 
the  some experimental observations of negligible inter-layer spin interactions~\cite{co8,co9}. 
The limited magnetic correlations between Co honeycomb layers along $c$ axis observed in the experiments on \ncto~was attributed to disordered distribution of 
Na$^+$ ions which disrupts any out-of-plane magnetic coupling making it essentially intra-layer.
Not to a complete surprise, the 3NN in \ncto~are also non-zero, similar to the cases of 
iridates and RuCl$_3$~\cite{mod11,mod12}.

Surprisingly, however, antiferromagnetic $J_3$ is larger in magnitude than $K_1$, hence being  the dominant interaction in \ncto. At first glance, it seems unusual given that  3$d$ orbitals are more compact compared to their 4$d$/5$d$ counterparts. In \ncto, the 3NN hopping can be mediated via Te ions sitting at the center of a Co-hexagon.  Orbital projected density of states for Te-$s$/$p$ orbitals (not shown) precludes any significant contribution near the Fermi-level 
where Co-$d$ and O-$p$ have their main contributions. This rules out any direct interaction between Co-$d$ and Te-$s/p$ orbitals. A second possibility for 
the 3NN interaction is through shared O-$p$ orbitals between these ions, which is favored by the highly suppressed  3NN
 Kitaev interaction due to a cancellation of  anistropic interactions from multiple long-distance pathways.  This then allows $J_3$ to become the largest  
magnetic interaction in this compound. Off-diagonal couplings of a smaller magnitude appears on these bonds. On a Z-bond, $\eta_3$ is found to be antiferromagnetic, while $\eta'_3$ ferromagnetic. However, on X or Y bond, signs of these couplings do not follow a specific pattern. On these two bonds, $\eta'_3$ never becomes ferromagnetic while $\eta_3$ can be either ferro- or 
 antiferro-magnetic. 

\begin{figure}[ht]
\centering 
\includegraphics[width=7.cm]{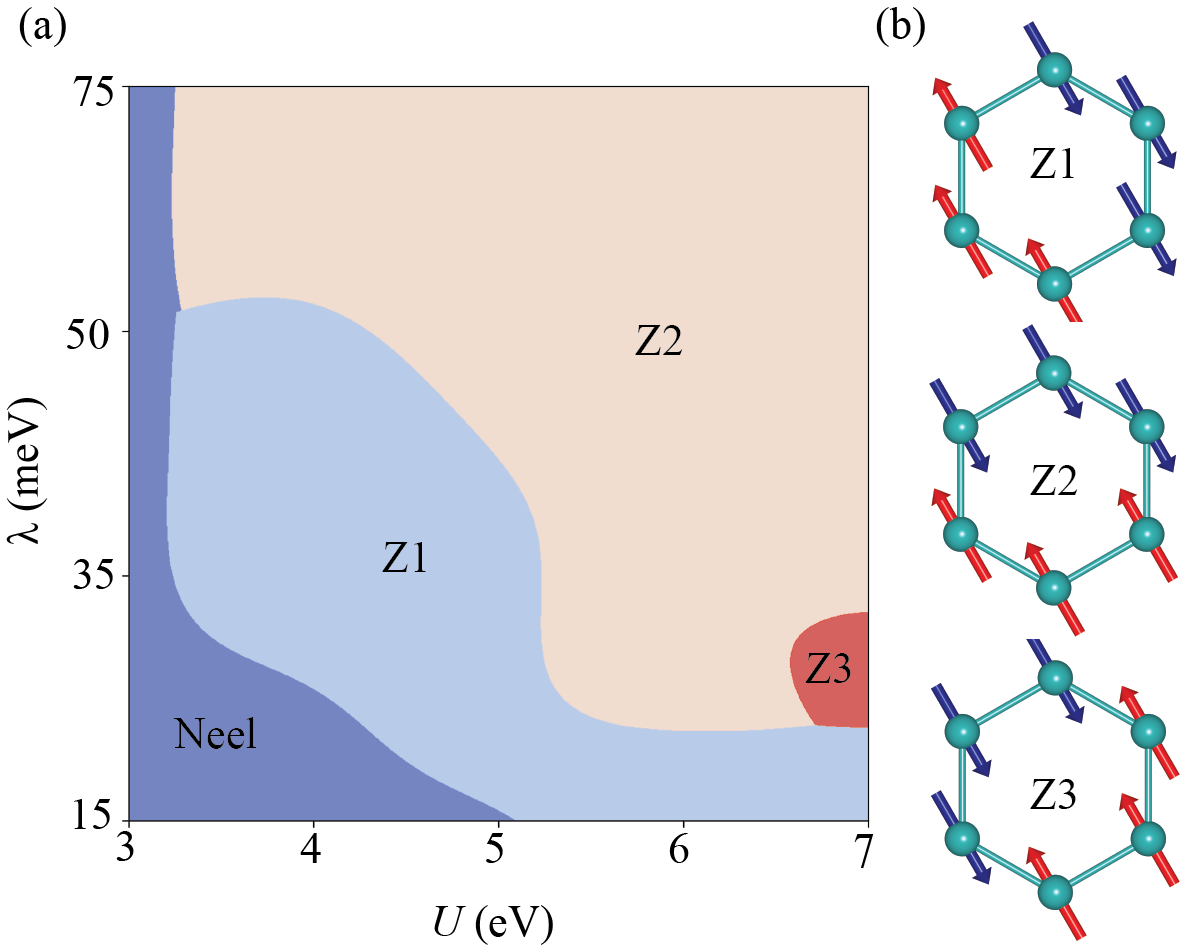}
\caption{(a) The $U$-$\lambda$ phase diagram for \ncto~ at $J_{\text{H}}$/$U$ = 0.16. Magnetic ground state obtained classically with magnetic interactions extracted  
in different regions of the phase space brings four distinct phases, a Neel and three zigzag phases Z1, Z2, Z3 shown in (b). These three zigzag phases differ by the direction of antiferromagnetic coupling 
 between the two ferromagnetic chains in a hexagon. Z1 is the experimentally proposed magnetic state in which antiferromagnetic coupling is along the X-bond 
 shown in Fig.~\ref{fig:fig03}(c).}
  \label{fig:fig04}
\end{figure}

 In order to examine whether the obtained parameters can reproduce the magnetic ground state of \ncto, we optimize the magnetic structure 
 using Hamiltonian of Eq.~(\ref{eqn:jk}) with spins allowed to rotate in the $a$-$b$ plane.
We find that the classical ground state was zigzag-type with the ferromagnetic chains running 
along $a$-direction coupled antiferromagnetically in direction perpendicular to $a$ (see Fig.~\ref{fig:fig04}(b)) with propagation vector 
$\sim$ (0.0, 0.5, 0.0). We call it Z2 structure. This structure is different than the experimentally proposed magnetic structure with propagation vector (0.5, 0.0, 0.0) which we call Z1. However, in the next section 
we show that one can also obtain Z1 as the classical 
ground by tuning the parameters of perturbation theory.

\subsubsection{Phase diagram}	

We varied $U$ and SOC strength ($\lambda$) to study their effect on magnetic interactions and on the ground state. 
A phase diagram, obtained from optimization of classical ground with magnetic interactions obtained at different $U$-$\lambda$ values,
is shown in Fig.~\ref{fig:fig04}(a).  $J_\text{H}/U$ ratio has been kept fixed at 0.16 in all these calculations. We obtain four distinct phase
viz, Neel and three types of zigzag magnetic order (Fig.~\ref{fig:fig04}(b)) differentiated by direction of antiferromagnetic coupling between the ferromagnetic chains. Phase diagram 
can be read in  the following manner.

In a smaller $U$ range (close to 3 eV), classical magnetic ground state is Neel-type and independent of $\lambda$. In this region,
 antiferromagnetic $J_1$ is dominant over other  types of first neighbour interactions. 
All these interactions increases with increase in $\lambda$ with  $K_1$ more rapidly than $J_1$. 
  At a particular $\lambda$, with the increase in $U$, we find decreasing $J_1$ and $K_1$ with $J_1$ decreasing more rapidly than $K_1$. 
  After critical values of $U$ and $\lambda$, situated on the phase boundary of Neel-Z1/Z2 phase,  $K_1$ term stars to the compete with $J_1$. Only when 
  $K_1$ $\ge$ $J_1$, the magnetic ground state changes from Neel to  one of the zigzag types. Z1 configuration (Fig.~\ref{fig:fig04}(b)) with propagation 
  vector $\sim$ (0.5, 0.0, 0.0) is the experimentally observed magnetic state of \ncto. In Z3 configuration 
 with propagation vector $\sim$ (0.5, 0.5, 0.0), antiferromagnetic coupling between ferromagnetic zigzag layers is along (1 1 0) direction. At some of the phase 
  points, spin moments of our classical ground state was found to slightly deviate from $b$-direction owing to small 
  $\eta$/$\eta'$ terms.

Here, a couple of remarks are in order. First, neither $J_1$, $K_1$ terms nor $J_3$, $\eta_3/\eta'_3$ terms alone can produce the zigzag ground state of 
\ncto~which is a collective efforts of all these terms. Second, these three zigzag spin configurations 
are distinct and hence break the three fold rotational symmetry. To verify this point we compared the energies of these three configurations at same phase point,
 $U$ = 4 eV  and $\lambda$ = 0.60 eV, the phase boundary point between Z1 and Z2 configurations. At this point, the obtained 
magnetic ground state was Z1 type and energies of Z2 and Z3 configurations were higher in energy by 0.010 meV and 0.015 meV per spin. 
These differences, when compared with the scale of magnetic coupling strengths, are significant and hence establish the fact of these three zigzag magnetic configurations being distinct. 
Our three zigzag configurations differ from those proposed in Chen {\it et.al.}~\cite{co9} as in the latter, the three domains of zigzag configurations are related by three fold rotational
 symmetry. For further validation of our obtained magnetic interactions,  spin-wave spectra are computed and analyzed next.

\subsection{Spin wave spectra}

\begin{figure}[ht]
  \centering
  \includegraphics[width=7.cm]{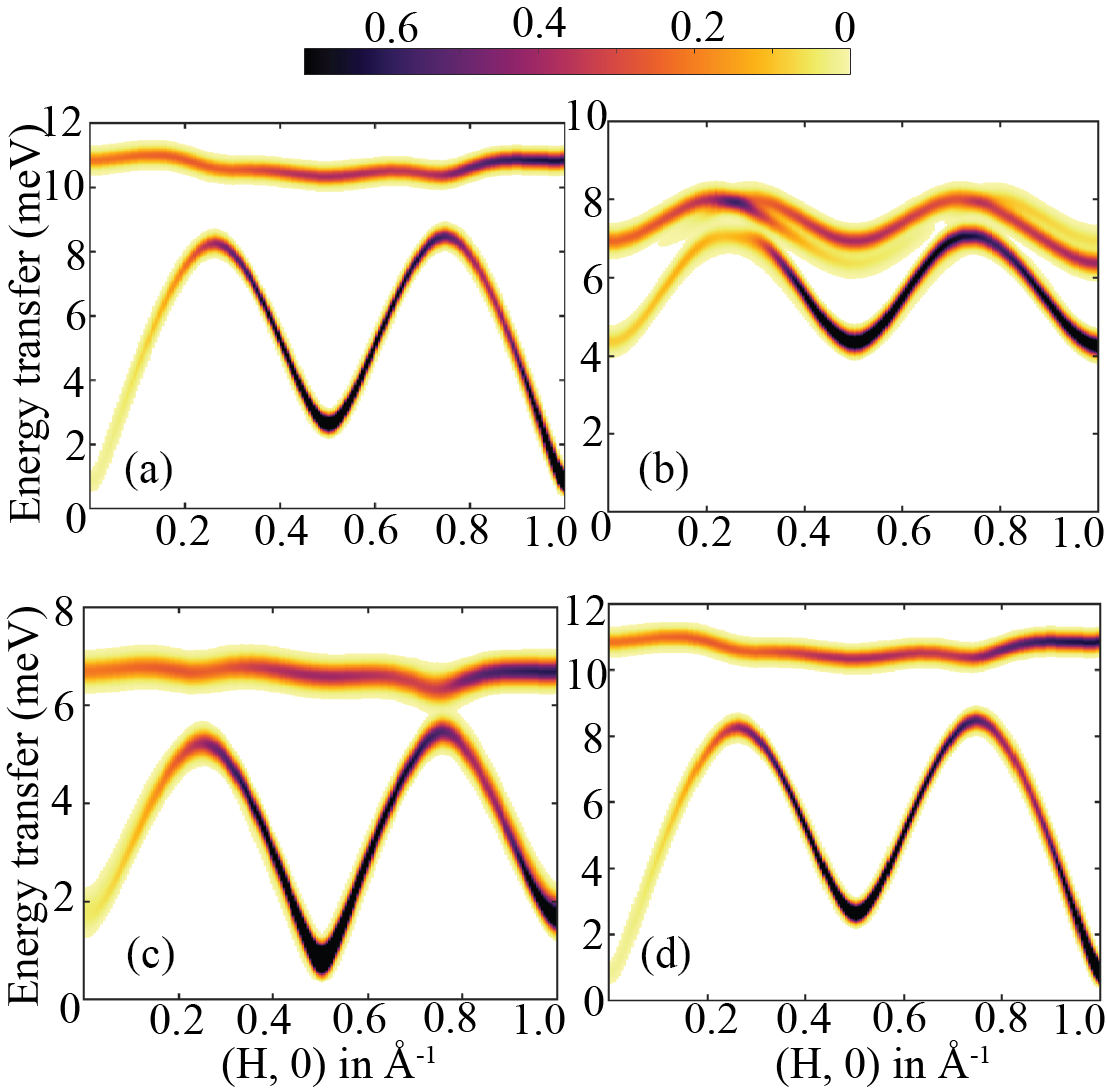}
  \caption{Spin wave spectra obtained from linear spin wave theory for the four phases shown in Fig.~\ref{fig:fig04}(a) namely, 
  (a) Neel at $U$ = 3.25 eV, $\lambda$ = 0.065 eV, 
  (b) Z1 at $U$- 4 eV, $\lambda$ = 0.06 eV, (c) Z2 at $U$ = 5 eV, $\lambda$ = 0.065 eV and (d) Z3 at $U$ = 7 eV, $\lambda$ = 0.025 eV.
  }
    \label{fig:fig05}
  \end{figure}

Recently, several inelastic neutron scattering experiments on \ncto~have reported spin wave spectra of this material~\cite{co3,co9}. 
The main feature of these reported spectra is gapped dispersive modes with lowest dispersive branch of width around  3 meV and some flat modes 
around 4.5 meV and 7 meV. A natural question about how well our estimated magnetic interactions can reproduce these experimental  findings is inevitable. 
The spin Hamiltonian in Eq.~(\ref{eqn:jk}) with estimated magnetic interactions is solved using the linear spin wave theory to obtain the spin wave spectra as 
implemented in SpinW package~\cite{spinw}.  
These calculations were performed 
for the four  distinct magnetic phases in  Fig.~\ref{fig:fig04}(a), and are shown in 
Fig.~\ref{fig:fig05}. 

The spectra of N\'eel, Z1 and Z3 states were found to differ from the experimentally obtained ones. 
First, dispersion width of the lowest branch in the spectra of N\'eel and Z3 phases (Fig.~\ref{fig:fig05}(a) and (d)) is much larger ($\sim$ 8 meV) than the experimentally observed $\sim$ 3 meV. 
Though this experimental feature of the spectra is captured in the case of Z1 phase (Fig.~\ref{fig:fig05}(b)), a gap of $\sim$ 1 meV with next higher dispersive mode observed in the experiments
 is missing in this case. 
Also, this gap is much larger (more than 2 meV) in our calculated spectra for Neel 
and Z3 phases. Therefore,  our calculated spin wave spectra for the Z2 phase shown in Fig.~\ref{fig:fig05}(c) closely resembles the experimental ones. Though, the band width of lowest 
branch is slightly higher ($\sim$ 5 meV) in our model, the computed spectra shown in (c) closely resemble the experimentally observed ones with an M-shaped lowest branch from Chen {\it et al}~\cite{co9} and the higher flat branch with a gap of 1 meV  
from Songvilay {\it et al}~\cite{co3}.
 Our calculated spectra also reproduces the gap of $\sim$ 1 meV with next higher branch, consistent with experimental findings~\cite{co3}.

\section{\texorpdfstring{\ncso}{TEXT}}
\label{ncso}

\begin{figure}[ht]
  \centering
  \includegraphics[width=7.cm]{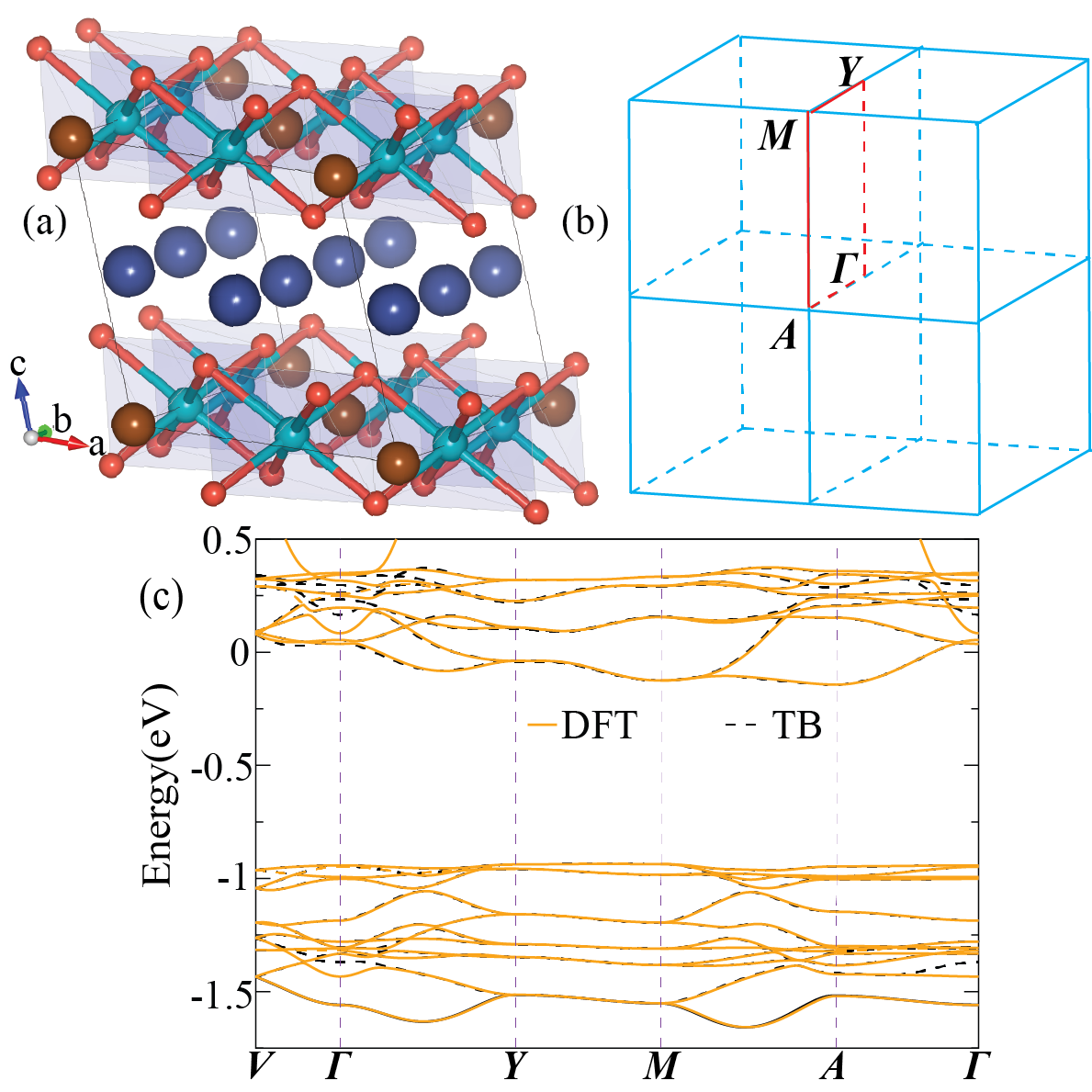}
  \caption{(a) Crystal structure of \ncso. Cyan, red, blue and brown spheres represent Co, O, Na and Sb atoms respectively. 
  Separation of Co-Sb planes along $c$ axis by Na layer is evident and edge shared Co-O octahedron are also shown. 
  (b) The Brillouin zone of \ncso, with some of the high-symmetry 
  points and paths. (C) TB band structure obtained by Wannierizing the $ab$ $initio$ electronic structure of 
  \ncso~considering all five Co-$d$ orbitals in the basis. Fermi energy is set to zero.}
    \label{fig:fig06}
  \end{figure}

The crystal of \ncso~is shown Fig.~\ref*{fig:fig06}(a). As for \ncso, Co-O-Sb environment is similar to that of Co-O-Te in \ncto, except that the 2-dimensional 
sheets of Co-O-Sb are joined by a triangular lattice of 
Na atoms stacking into the three-dimensional lattice along $c$ direction with space group $C2/m$. Thus, similar to \ncto, 
the magnetic lattice comprised of a Co-honeycomb network in \ncso~is 
quasi-two dimensional and magnetic ground state is again zigzag type antiferromagnet described by propagation vector (1/2, 1/2, 0) with the spins aligned along the 
crystallographic $b$ axis. However, Co-Co distances in \ncso~are 10\% shorter than \ncto. It would hence be interesting to investigate whether such differences in structural features 
can affect magnetic interactions in \ncso. The Brillouin zone of \ncso~is shown in Fig.~\ref{fig:fig06}(b).
Following similar computational procedures to that of \ncto, we extracted the CF matrix and hopping interactions for \ncso~and fitting of  
$ab$ $initio$ band structure to a Wannier TB model is shown in Fig.~\ref{fig:fig06}(c). 

The obtained 
CF matrix and hopping amplitudes are given in Eq.~\ref{eq:cf2} and~\ref{appendix:hop_ncso}, respectively. Inspection of O-Co-O angles in an octahedra 
reveal a deviation from ideal case of 90$^o$ to the largest angle being $\sim$ 96$^o$ and smallest one is $\sim$ 81$^o$. 
Thus, the lower symmetry of $C2/m$ space group along 
with the trigonal CF causes the 
additional splitting of 15 meV within $e_g$ orbitals and two types of splitting within 6 meV, 12 meV splitting within the $t_{2g}$ orbitals. 
This is consistent with the previous findings on Iridates~\cite{mod11} and RuCl$_3$~\cite{mod12}. 
Co-O-Co angles across a pair of edge-sharing octahedra are $\sim$ 93.5$^o$, again close to 90$^o$ but slightly larger 
than the case of \ncto.  
In our ``onsite'' Hamiltonian for \ncso, $U$ and $\lambda$ were kept the same to 5 eV and 65 meV as before. Similar to the case of \ncto, in this case too, the lowest six states are three  Kramer's doublet with energy 
separation of 28 and $\sim$ 13 meV, between the lowest and the higher two doublets, respectively.

\begin{equation}
\label{eq:cf2}
\Delta_i = 
\begin{bmatrix*}[r]
1.344  & 0.133 & -0.035 & 0.002 &  -0.084 \\
0.133 &  0.046 & 0.007 & -0.180  & 0.020 \\	
-0.035 &  0.007 & 0.016 &  -0.093  & 0.022 \\
0.002 & -0.180 & -0.093 &  1.309 &  -0.224 \\
-0.084 & 0.020 & 0.022 &  -0.224 &  0.045 \\
\end{bmatrix*}
\end{equation}

\begin{figure}[ht]
  \centering
  \includegraphics[width=7.cm]{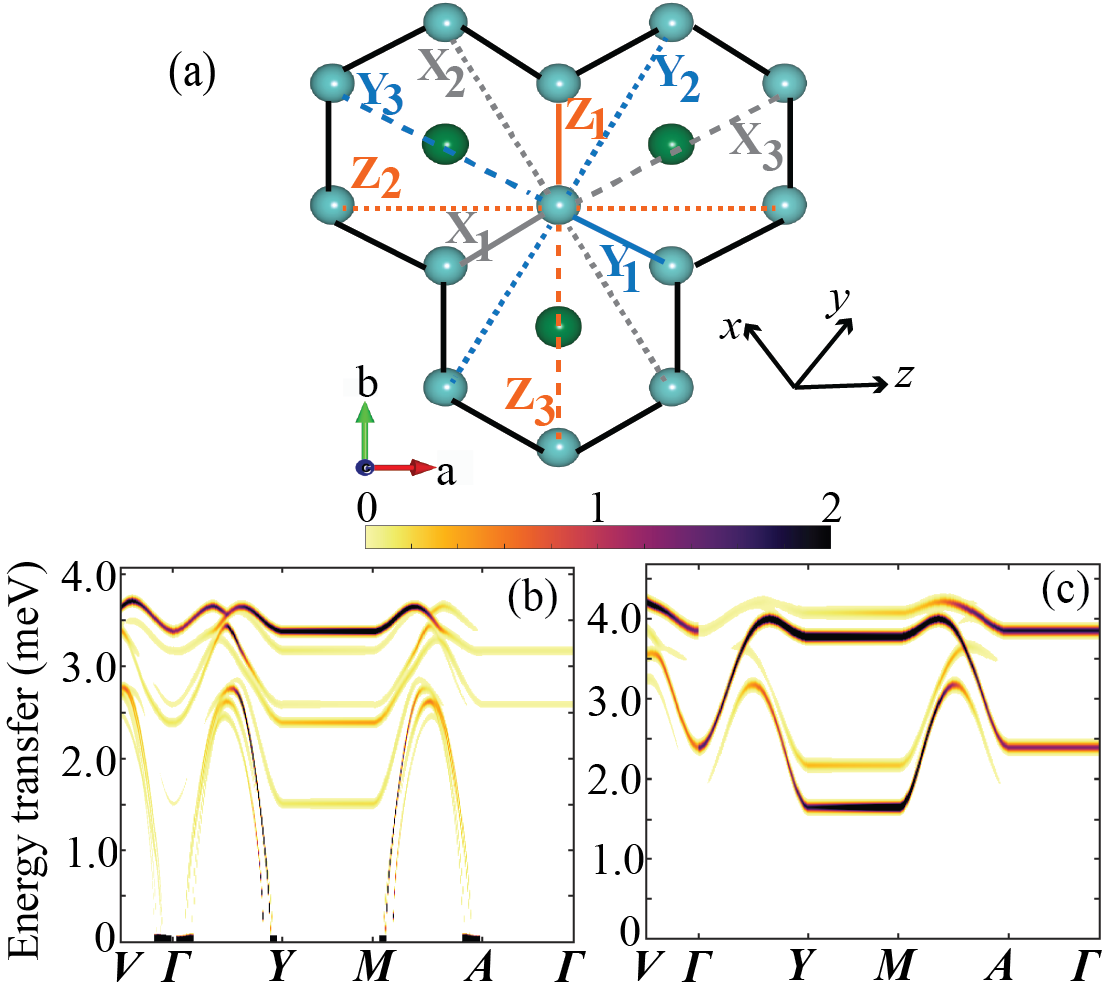}
  \caption{(a) Lattice of Co$^{+2}$ in \ncso. 1NN, 2NN and 3NN 
   are shown by solid, dotted and dashed lines. Orange, dark gray and skyblue  color represent the Z, X and Y bonds respectively. $a$, $b$ and $c$ are the crystallographic axes and cubic 
   axes of Co-O octahedra are denoted by $x$, $y$ and $z$. (b) Spin wave spectra obtained from linear spin wave theory for optimized classical ground state of 
   \ncso~(at $U$ = 5.0 eV, $\lambda$ = 0.065 eV), 
   for the case when only in-plane movements of spins were allowed during the optimization. (c) Spectra obtained after full classical optimization of magnetic ground state for which a large out-of-plane deviation
    of magnetic moments 
   is observed while zigzag configuration remains the ground state.}
    \label{fig:fig07}
  \end{figure}

For the lower symmetry space group of \ncso, symmetry-inequivalent nearest-neighbour bonds have two types, $viz$-a-$viz$, Z$_1$ bonds 
$\parallel$ to 
crystallographic $b$ axis with  a local $C_{2h}$ point group symmetry and X$_1$/Y$_1$ bonds lying in the $ab$ plane with a (lower) $C_i$ point group 
symmetry. This arrangement of bonds is shown in Fig.~\ref{fig:fig07}(a). In such a case, on these X/Y type bonds, 
in addition to $J$, $K$, $\Gamma$ and $\Gamma'$, one needs additional 
parameters 
$\zeta$ and $\xi$ as explained in Winter {\it et. al.}\cite{mod11} to describe the magnetic interactions. 
The magnetic interactions estimated in this case using the second-order perturbation theory is listed in Table~\ref{tab:ncso}. 

\begin{table}[!ht]
  \centering
 \begin{tabular}{lSSSSSSS}
 \hline
  
   Bond type   & $J$   & $K$ & $\eta$ & $\eta'$ & $\xi$ & $\zeta$  \\
 \hline \\
 Z$_1$ &  0.3    & -0.757  &   0.0 & 0.0 & 0.0 &  0.0    \\ 
 
 X$_1$/Y$_1$ &  0.186   & -0.211  &   0.0 & -0.15 & -0.421 &   0.0   \\ 
 
 Z$_3$ &  1.7    & 0.4  &   0.34 & -0.31 & 0.0 &  0.0    \\ 
 
 X$_3$/Y$_3$ &  1.823   & -0.189  &   0.31 & 0.0 & -0.211 &   0.302	   \\  \\
 \hline
 \end{tabular}
 \caption{ Estimated first and third NN Heisenberg $J$, Kitaev $K$ and off-diagonal $\eta$,$\eta'$ terms with additional 
 parameters $\xi$ and $\zeta$ in \ncso~given in meV. The 2NN interactions were found to be negligibly small. 
 Parameters used are $U$ = 5 eV, 
 $J_{\text{H}}$ = 0.8 eV and $\lambda$ = 65 meV.}
 \label{tab:ncso}
 \end{table}

By comparing with \ncto, we find that the 1NN magnetic interactions 
are more or less similar in this case. However, a subtle difference can be found for the 3NN interactions as in \ncso, the magnitudes are half 
of their values of \ncto. 
One more difference between these two materials is that the Kitaev coupling on Z$_3$ bond, though again smaller, is found to be antiferromagnetic in this case. 
Additionally, the $\eta_3$/$\eta'_3$ terms are much smaller (almost half) in this case when compared to \ncto.

With these magnetic parameters, in optimization of classical ground state using the Hamiltonian in Eq.~\ref{eqn:jk}, only in-plane movement of spins was allowed first which
 brings the same zigzag type 
magnetic ground which was obtained experimentally. However, we find a small deviation of spin from $b$ direction along $a$-axis
 in our optimized structure in this case too. In terms of propagation vector, the optimized magnetic structure is described by $\sim$ (0.498, 0.492, 0). Using this optimized magnetic
  structure we calculated the spin wave spectra 
 of \ncso~shown in Fig.~\ref{fig:fig07}(b). Most of the intensity of this spectra is found in 3-4 meV energy range which is in reasonably good 
 agreement with the recent experiment~\cite{co3}. However, the negative magnon frequencies imply that the optimized magnetic ground state may not be fully consistent with 
 the magnetic interactions of Table~\ref{tab:ncso}. A full classical optimization of magnetic structure again leads to a zigzag ground state, in which spins show significant out-of-plane tilting. Spectra obtained with this structure is shown in Fig.~\ref{fig:fig07}(c), 
 which clearly is free of negative frequencies and and show a 2 meV Goldstone gap. 
  This gapped spectral feature is consistent with the previous experimental finding~\cite{co3}.

 We now will briefly compare the properties of these two materials.  Structurally, the crucial difference is the stacking of the 
 Co honeycomb layers. For \ncto, the layers are staggered while for \ncso, these layers are exactly on the top of each other. However, 
the estimated interlayer magnetic interactions in our calculations are found to be negligibly small ruling out any possibility 
of coupling between the honeycomb layers of Co in both these materials. Co-Co distance in these two materials is very similar and comparatively 
larger deviation of Co-O-O angle from 90$^o$ between edge shared octahedra is found in \ncso~($\sim$ 93.5$^o$) when compared to the 
\ncto~(92$^{o}$)~\cite{co5} as discussed earlier. This small difference does not seem to substantially affect the nearest-neighbour magnetic interactions. 
So, while the 1NN interactions have similar strengths in both the materials (after averaging of two type of bonds in \ncso), the smaller 3NN magnetic interactions in \ncso~can be the reason behind its lower T$_N$ ($\sim$ 7 K) than \ncto~ (T$_N$ $\sim$ 18 K). 
Also, the smaller off-diagonal $\eta_3$/$\eta'_3$ in \ncso~ can be seen as 
a manifestation of the lower trigonal CF in \ncso~as been explained in Liu {\it et. al.}~\cite{co4}. Thus, the smaller ``undesirable'' 
off-diagonal and 3NN terms make \ncso~ a more suitable candidate than \ncto~in the quest of spin-liquid behaviour.

In a recent study by Das \textit{et. al.}~\cite{shreya} on cobalt-based Kitaev materials, the authors commented, en passant, on the magnetic interactions in \ncso. They found 1NN ferromagnetic 
Heisenberg and off-diagonal, 
antiferromagnetic 1NN Kitaev and 3NN Heisenberg couplings, all are of equal strength for this compound, differing from our results. It is important to verify that the combination of their 
magnetic  interactions  stabilizes the zigzag ground state, from following two considerations. First, the consensus seems to be  that the zigzag ground state in these materials is a result
 of a combination of ferromagnetic $K_1$ and antiferromagtic  $J_1$ and $J_3$~\cite{co1,co4}. 
Second, though with 1NN antiferro Kitaev coupling where the nature of Heisenberg term is of lesser consequence, one may also realize a zigzag ground state. 
However, in this case the role of longer-range Heisenberg interaction requires further clarification.~\cite{co1,co4}

\section{Summary and remarks}

In this paper, we investigated the structural and magnetic properties of \ncto~and \ncso. Using a cluster expansion model, we find a particular 
pattern of hexagonal lattice formation by Na-vacancies in \ncto. Using a multiband Hubbard model and the second-order perturbation theory valid in the large $U$ regime,  we have estimated the magnetic interactions in these material and showed that a zigzag magnetic order can only be stabilized when Kitaev term 
is either comparable to or larger than the first neighbour Heisenberg term. However, we find the third neighbour Heisenberg interaction to be the 
dominant magnetic interactions in both these compounds.
In addition, we have also presented a phase diagram for 
\ncto ~by varying $U$ and $\lambda$, a majority portion of which  is shared by different zigzag magnetic ground state. Based on the spin Hamiltonian, we have calculated spin wave spectra
 and showed that our findings are in qualitative agreement with the recent neutron-scattering experiments.
A comparison between these two materials establish the fact that \ncso~is might be a better candidate than \ncto~ in the quest for Kitaev materials due to smaller off-diagonal and 3NN magnetic 
interactions in the former.

The trigonal CF present in these materials is believed to be the main hurdle for realization of a quantum spin liquid phase and has been recently suggested as an experimentally tuneable
 parameter~\cite{co4}. 
Previously, this type of tuning has been achieved experimentally by means of strain in CoO~\cite{press_tri} and would be an interesting future direction of research at both theoretical and 
experimental fronts. 
For another interesting aspect from an experimental perspective, it will be worth investigating whether magnetic field can induce a quantum spin liquid 
phase in these cobalt-based compounds, alluding to another Kitaev candidate material $\alpha$-RuCl$_3$~\cite{qsl_mag_fld}, .

\begin{acknowledgments}
We are grateful for stimulating discussions with Yuan Li. We acknowledge the financial support from the National Natural Science Foundation of China (Grants No. 11725415 and 11934001), the National Key R\&D Program of China (Grants No.2018YFA0305601 and 2021YFA1400100), and the Strategic Priority Research Program of Chinese Academy of Sciences (Grant No. XDB28000000).
\end{acknowledgments}

\newpage
\appendix
\renewcommand{\thesection}{Appendix A}

\begin{widetext}
\section{1NN and 3NN hopping  amplitudes in \texorpdfstring{\ncto}{TEXT} }
\label{appendix:hop}

\begin{table}[ht]
 \centering
\begin{tabular}{lSSSSSSS}
\hline
  &\multicolumn{3}{c}{1$^{st}$ NN} & & \multicolumn{3}{c}{3$^{rd}$ NN}\\
\cline{2-4} \cline{6-8}
  Hopping   & X   & Y & Z & & X & Y & Z\\
\hline
 $d_{z^2} \rightarrow d_{z^2}$ &  -0.0307    &  -0.0307 &   -0.0709 & & 0.0806 & 0.0806 & -0.0335 \\
  $d_{zx} \rightarrow d_{zx}$ &  0.0364       &  -0.1699      &  0.0364 & & 0.0047 & -0.0327 & 0.0047  \\
$d_{yz} \rightarrow d_{yz}$ &   -0.1699    &  0.0364      & 0.0364 & & -0.0327 & 0.0047 & 0.0047 \\
 $d_{x^2-y^2} \rightarrow d_{x^2-y^2}$ &  -0.0575    & -0.0575     & 0.0046 & & 0.0046 & 0.0046  & 0.1187 \\
 $d_{xy} \rightarrow d_{xy}$ &    0.0364   & 0.0364       & -0.1699 & & 0.0047 & 0.0047 & -0.0327 \\
 $d_{z^2} \leftrightarrow d_{zx}$ &  -0.0184    & 0.0638  &   0.0071 & & 0.0082 & 0.0080 & -0.0010 \\
 $d_{z^2} \leftrightarrow d_{yz}$ &   0.0638   & -0.0184  &   0.0071 &  & 0.0085 &0.0073  & -0.0010 \\
 $d_{z^2} \leftrightarrow d_{x^2-y^2}$ &  0.0232    & -0.0232  &   0.0000 & & 0.0659 & -0.0659 & 0.0000 \\
 $d_{z^2} \leftrightarrow d_{xy}$ &  -0.0114    & -0.0114  &   0.1275 & & 0.0072 & 0.0063  & 0.0161 \\
$d_{zx} \leftrightarrow d_{yz}$ &   0.0467     & 0.0467      &  -0.0457 & & -0.0007 & -0.0007  & -0.0077 \\ 
$d_{zx} \leftrightarrow d_{x^2-y^2}$ &  -0.0030       & 0.1104       & -0.0176 &&  0.0030 & 0.0146 & 0.0089  \\
$d_{zx} \leftrightarrow d_{xy}$ &    0.0457     & -0.0467      &  -0.0467 & & 0.0077 & 0.0007 & 0.0006 \\
$d_{yz} \leftrightarrow d_{x^2-y^2}$ &   -0.1098    & 0.0030       & 0.0176 & &  -0.0139 & -0.0036 & -0.0089 \\
$d_{yz} \leftrightarrow d_{xy}$ &   -0.0479    & 0.0457       & -0.0467 & & 0.0006 & 0.0077 & 0.0006\\
$d_{x^2-y^2} \leftrightarrow d_{xy}$ &  -0.0147   & 0.0146 &   0.0000  & & 0.0053  & -0.0048 & 0.0000 \\
\hline 
\end{tabular}
\end{table}

\renewcommand{\thesection}{Appendix B}
\section{1NN and 3NN hopping  amplitudes in \texorpdfstring{\ncso}{TEXT} }
\label{appendix:hop_ncso}

\begin{table}[ht]
 \centering
\begin{tabular}{lSSSSSSS}
\hline
  &\multicolumn{3}{c}{1$^{st}$ NN} & & \multicolumn{3}{c}{3$^{rd}$ NN}\\
\cline{2-4} \cline{6-8}
  Hopping   & X   & Y & Z & & X & Y & Z\\
\hline
 $d_{z^2} \rightarrow d_{z^2}$ &  0.0001    &  -0.0096 &   -0.0590 & & 0.0060 & 0.0011 & 0.0244 \\
  $d_{zx} \rightarrow d_{zx}$ &  -0.1478       &  0.0390      &  0.0352 & & 0.0049 & 0.0073 & -0.0010  \\
$d_{yz} \rightarrow d_{yz}$ &   0.0483    &  -0.1449      & 0.0272 & & 0.0055 & 0.0047 & -0.0071 \\
 $d_{x^2-y^2} \rightarrow d_{x^2-y^2}$ &  -0.0624    & -0.0463     & 0.0060 & & 0.0060 & 0.0206  & 0.0018 \\
 $d_{xy} \rightarrow d_{xy}$ &    0.0365   & 0.0365       & -0.1312 & & 0.0009 & -0.0103 & 0.0029 \\
 $d_{z^2} \leftrightarrow d_{zx}$ &  -0.0482    & 0.0074  &   -0.0219 & & -0.0124 & 0.0176 & -0.0159 \\
 $d_{z^2} \leftrightarrow d_{yz}$ &   -0.0021   & 0.0616  &   0.0220 &  & -0.0158 & 0.0089  & 0.0096 \\
 $d_{z^2} \leftrightarrow d_{x^2-y^2}$ &  -0.0167    & 0.0270  &   -0.0223 & & 0.0175 & -0.0150 & 0.0016 \\
 $d_{z^2} \leftrightarrow d_{xy}$ &  0.0142    & 0.0059  &   -0.1213 & & -0.0018 & 0.0016  & -0.0167 \\
$d_{zx} \leftrightarrow d_{yz}$ &   -0.0106    & -0.0336      &  0.0234 & & 0.0059 & -0.0019  & 0.0147 \\ 
$d_{zx} \leftrightarrow d_{x^2-y^2}$ &  -0.1229       & 0.0032       & 0.0057 &&  -0.0167 & -0.0061 & 0.0134  \\
$d_{zx} \leftrightarrow d_{xy}$ &    -0.0094     & 0.0239      &  -0.0497 & & 0.0030 & 0.0042 & -0.0094 \\
$d_{yz} \leftrightarrow d_{x^2-y^2}$ &   0.0079    & -0.1114    & 0.0205 & &  -0.0154 & -0.0171 & 0.0134 \\
$d_{yz} \leftrightarrow d_{xy}$ &   -0.0221    & 0.0359       & 0.0529 & & -0.0423 & -0.0026 & -0.0060\\
$d_{x^2-y^2} \leftrightarrow d_{xy}$ &  0.0066   & 0.0189 &   -0.0246  & & -0.0278  & 0.0229 & 0.0013 \\

\hline 
\end{tabular}
\end{table}
\newpage
\end{widetext}


%

\end{document}